\documentclass[twoside]{article}
\usepackage{fleqn,espcrc2}

\newcommand {\be}{\begin{equation}}
\newcommand {\ee}{\end{equation}}
\newcommand {\beq}{\begin{eqnarray}}
\newcommand {\eeq}{\end{eqnarray}}
\newcommand {\pr}{\partial}
\newcommand {\lf}{\lefteqn}
\newcommand {\sg}{\sigma}
\newcommand {\sgm}{\sigma^{'}}
\newcommand {\al}{\alpha}
\newcommand {\alp}{\alpha^{'}}
\newcommand {\dl}{\delta}
\newcommand {\ep}{\epsilon}
\newcommand {\non}{\nonumber}

\newcommand{\AmS}{{\protect\the\textfont2
  A\kern-.1667em\lower.5ex\hbox{M}\kern-.125emS}}

\title{Bihamiltonity as origin of T-duality of the closed string model}

\author{V. D. Gershun\thanks{e-mail: gershun@kipt.kharkov.ua} \address
{Institute of Theoretical
Physics, NSC Kharkov Institute of Physics and Technology, \\ P.O.
Box 310108, 1 Akademicheskaya St., Kharkov,
Ukraine}}

\begin{document}
\begin{abstract}
 In assumption,that string model is the integrable model for particular
kind of the background fields, the closed string model in the background
gravity field and the antisymmetric  B-field is considered as the
bihamiltonian system. It is shown, that bihamiltonity is origin of
T-duality of the string models. The new Poisson brackets,
depending of the background fields and of their derivatives, are
obtained. The integrability condition is obtained as the
compatibility of the bihamiltonity condition and the Jacobi
identity of the new Poisson bracket.  The B-chiral string model
is dual to the chiral string model for the constant background
fields.

\vspace{1pc}
\end{abstract}

\maketitle
\section{Introduction}
 The first kind constraints in a gauge theory are the generators of
 the gauge transformations. All of the Fourier components of constraints
are the integrals of motion and does not zero component only, on
the constraints surface. Then, algebra of the integrals of motion
can be non Abelian. Two dynamical systems are dual, if they are
described by their sets of the integrals of motion $H_{i}, F^{k}$
the same dimension, which are in involution between themselves.
\be\lf{\{H_{i}, H_{k}\}=0, \{F^{a}, F^{b}\}=0,\,\,i,a=1...N}\ee
bihamiltonian system \cite{ft,mbps,od} is dynamical system with N
integral of motions $H_{1},...H_{N}$ in involution, if any of the
integrals of motion can be considered as hamiltonian, and the
following condition is be satisfied \be\lf{\dot
x^{a}=\{x^{a},H_{1}\}_{1}=...=\{x^{a},H_{N}\}_{N}}\ee Also, the
new hierarchies of the Poisson brackets (PB) are arised. It is
more interesting, if the new hierarchies of the equation of
motion arise under the new time coordinates $t_{k}$.
\be\lf{\frac{dx^{a}}{dt_{n+k}}=\{x^{a},H_{n}\}_{k+1}}\ee
where $n,k=1...N$. The new equations of motion are describe the new
dynamical systems with the same set of the integrals of motion, which are
dual to the original system. The dual set of the integrals of motion can be
obtained from the original it by the mirror transformations and by the
contraction of the constraints algebra. The contraction of the constraints
algebra means, that dynamical system is belong to the orbits of
corresponding generators and is describe the invariant subspace. The set
of involutive integrals of motion is belong to Cartan subalgebra of the
constraints algebra.  Consequetly, duality is property of the integrable
models.  T-duality in integrable models have considered in Ref.
\cite{fgnr,g} by the separation of variables method. This duality is
correspond to the canonical transformations from one dynamical system to
other it. Duality in the string models is appeared in analisis of toroidal
compactifications \cite{bgs,ky}.  It was shown, that theories defined on
the circles of radii $R$ and $\frac{2}{R}$ are equivalent. In Ref.
\cite{b} was given general prescription to find dual $\sigma$- models on
the manifolds with Abelian isometries. T-duality of the
compactified string is result of the symmetry of constraints and
of the (PB)
\beq\lf{H(\tau,\sg)=\frac{1}{2}[2\pi\alp Rp_{a}^{2}+\frac{1}{2\pi \alp
R}x_{a}^{'2}]}\non \\ \lf{P(\tau,\sg)=p_{a}x_{a}^{'}}\\
\lf{\{x_{a}(\sg),p_{b}(\sgm)\}_{1}=\dl_{ab}\dl(\sg -\sgm)}\non\eeq
under transformations $p_{a}\to x_{a}^{'}$,
$2\pi\al^{'}R\to\frac{1}{2\pi\al^{'}R}$. The nonlocal coordinates \cite{oa}
\beq\lf{x_{a}(\sg)=\int\limits_{0}^{\sg}{p_{a}(\sgm)d\sgm}}\\
\lf{[x_{a}(\sg),x_{b}(\sgm)]\approx \theta(\sg -\sgm)}\non\eeq
arise. This kind of T-duality is described by canonical
transformations \cite{cz,agl1,agl2} and it is transform $H\to H$,
$P\to P$.  Thus, there are two essentially different kinds of
T-duality. First it is described by canonical transformations and
it is transform $H_{k}\to H_{k}$, $\{,\}_{k}\to \{,\}_{k}$.
Second it is described by noncanonical transformations \cite{ger}
and it is transform $H_{k}\to H_{l}$,
$\{,\}_{k}\to \{,\}_{l}$, $k\ne l$. We are suppose, that string model, in
the background gravity field and antisymmetric B-field, is the
integrable model for particular kind of the background fields and
we are use bihamiltonian approach to investigation of the
T-duality  and of the integrability of the closed string models.
Any of generators of $Vir\oplus Vir$ algebra can be considered as
hamiltonian. But, there is only one model, dual to original, due
to two elements of Cartan subalgebra. It is shown,that string
model is bihamiltonian system for free string and for constant
background fields. The free chiral string is dual string to free
original string. The B-chiral string in the constant background
fields is dual to chiral string also. In the case of the
arbitrary background fields, it is obtained the new (PB),
depending of the background fields and of their derivatives. The
integrability condition is obtained as the compatibility of the
bihamiltonity condition and of the Jacobi identity for the new
(PB).  \section{T-duality of the free closed string model} The
free closed string model in conformal gauge is described by
Lagrangian
\be\lf{L=\frac{1}{2}[\dot x_{a}^{2}+x_{a}^{'2}]}\ee
and by first kind constraints
\beq\lf{H_{1}=\frac{1}{2}[2\pi\alp p_{a}^{2}+\frac{x_{a}^{'2}}{2\pi\alp}]
\approx 0}\\
\lf{H_{2}=p_{a}x_{a}^{'}\approx 0,\,\,\,x_{a}(\sg
+2\pi)=x_{a}(\sg)}\non\eeq
These constraints form $Vir\oplus Vir$ algebra under (PB) $\{,\}_{1}$.
\beq\lf{\{L_{n},L_{m}\}_{1}=-i(n+m)L_{n-m}}\non \\
\lf{\{\bar L_{n},\bar L_{m}\}_{1}=-i(n-m)\bar L_{n-m}}\label{com}\non \\
\lf{\{L_{n},\bar L_{m}\}_{1}=0}\eeq
\beq\lf{L_{k}=\frac{1}{4\pi}\int\limits_{0}^{2\pi}(H_{1}+H_{2})
e^{ik\sg}d\sg}\\
\lf{\bar L_{k}=\frac{1}{4\pi}\int\limits_{0}^{2\pi}(H_{1}-H_{2})
e^{ik\sg}d\sg}\eeq
Hamiltonian equations of motion under hamiltonian $H_{1}$ are
\beq\lf{\dot
x_{a}(\sg)=\int\limits_{0}^{2\pi}d\sgm\{x_{a}(\sg ),H_{1}(\sgm )\}_{1}=
2\pi\alp p_{a}}\non \\
\lf{\dot p_{a}=\frac{1}{2\pi\alp }x_{a}^{''}}\eeq
The classical solution is
\beq\lf{x^{a}(\tau,\sg )=x^{a}(0)+\alp p^{a}\tau+}\label{eq}\\
\lf{\sqrt{\frac{\alp}{2}}\sum[\frac{c_{n}^{a}}{\sqrt{n}}e^{-in(\tau +\sg
)}+ \frac{\bar c_{n}^{a}}{\sqrt{n}}e^{-in(\tau -\sg
)}+c.c.]}\non\eeq The bihamiltonity conditions are modified
\beq\lf{\dot x_{a}=\{x_{a},\int\limits_{0}^{2\pi}d\sgm
H_{1}\}_{1}\cong
\{x_{a},\int\limits_{0}^{2\pi}d\sgm H_{\pm 2}\}_{\pm 2}}\non \\
\lf{= 2\pi\alp p_{a} - \alp P_{a}}\\
\lf{\dot
p_{a}=\frac{1}{2\pi\alp}x_{a}^{''},\,\,P_{a}=\int\limits_{0}^{2\pi}d\sg
p_{a}(\sg)=const.}\eeq
The dual (PB) are
\beq\lf{\{x_{a},x_{b}\}_{\pm 2}=\pm\dl _{ab}[\pi\alp \ep (\sgm -\sg)}\non
\\ \lf{-\alp (\sgm -\sg )]}\label{ss}\\ \lf{\{p_{a}(\sg ),p_{b}(\sgm
)\}_{\pm 2}=\pm 2\pi\alp \dl _{ab}\frac{\pr}{\pr \sg }\dl(\sg -\sgm )}\non
\\ \lf{\{x_{a}(\sg ),p_{b}(\sgm )\}_{\pm 2}=0}\eeq The (PB) (\ref{ss}),
without last term, was introduces in Ref. \cite{ws,js}. This term is
reduce to absence of the total momentum in (\ref{eq}). The $Vir\oplus Vir$
algebra has wrong sign on comparison with (\ref{com}) under (PB)
$\{,\}_{2}$. To satisfy the bihamiltonity condition, the following dual
map is necessary:  \be\lf{H_{1}\to \pm H_{2}, L_{0}\to \pm
L_{0},\bar L_{0}\to \pm \bar L_{0}, \tau\leftrightarrow \sg}\ee
The Gupta-Bleyler
 quantization of this model is reduce to the same mass spectrum and to the
same wave functions, with regard to the dual mapping.  The dual
dynamical system is defined by equations \beq\lf{\dot
x_{a}=\{x_{a},\pm\int\limits_{0}^{2\pi}d\sgm H_{2}\}_{1}=}\non \\
\lf{\{x_{a},\int\limits_{0}^{2\pi}d\sgm H_{1}\}_{2}=\pm x_{a}^{'},\, \dot
p_{a}=\pm p_{a}^{'}}\eeq
and is describe the left(right)chiral string
$x_{a}(\tau,\sg )=x_{a}(\tau \pm \sg )$. It means, that $L_{n}=0$, or
$\bar L_{n}=0$, $n\ne 0$. The Gupta-Bleyler quantization of the chiral
string is coincide with the quantization of the closed string with
additional condition of "freezing" modes $L_{n}=0$, or $\bar
L_{n}=0$,
$n\ne 0$ in the strong sense. Let me remember, that both dynamical systems,
the original string and the chiral string, is described by the same
Lagrangian.  Also, it is possible to consider hamiltonians $L_{k}+\bar
L_{k}$ together with (PB) \be\lf{\{x_{a}(\sg ),p_{b}(\sgm
 )\}_{k}=e^{ik\sgm }\dl _{ab}\dl (\sg - \sgm)}\ee and hamiltonians
$L_{k}-\bar L_{k}$ together with (PB) \beq\lf{\{x_{a}(\sg ),x_{b}(\sgm
)\}_{k}=\dl _{ab}[\pi\alp \ep (\sgm -\sg )-}\non \\ \lf{-\alp (\sgm -\sg
)]e^{ik\sgm}}\\ \lf{\{p_{a}(\sg ),p_{b}(\sgm )\}_{k}=\dl
_{ab}\frac{1}{2\pi\alp }e^{ik\sgm }\frac{\pr}{\pr \sg }\dl (\sg -\sgm
)}\non\eeq But, this hamiltonians do not generate the new dynamical
systems.

\section{Closed string in the background fields} The closed
 string in the background gravity field and the antisymmetric B-field is
described by first kind constraints
\beq\label{ham}\lf{H_{1}=\frac{1}{2}g^{ab}(x)[p_{a}-\al
B_{ac}(x)x^{'c}][p_{b}-\al B_{bd}(x)x^{'d}]}\non \\
\lf{+\frac{1}{2}g_{ab}(x)x^{'a}x^{'b},\,\, H_{2}=p_{a}x^{'a}}\eeq
where $a,b =0,1,...D-1$,$\al$ -arbitrary parameter. The original (PB) are
\beq\lf{\{x^{a}(\sg ),p_{b}(\sgm )\}_{1}=\dl _{b}^{a}\dl (\sg -\sgm )} \\
\lf{\{x^{a},x^{b}\}_{1}=\{p_{a},p_{b}\}_{1}=0}\non\eeq
The dual (PB) has following form
\beq\label{new}\lf{\{A(\sg ),B(\sgm )\}_{2}=}\\
\lf{\frac{\pr A}{\pr x^{a}}\frac{\pr B}{\pr x^{b}}[[\omega^{ab}(\sg
)+\omega^{ab}(\sgm )]\ep (\sgm -\sg )+}\non \\ \lf{[\Phi^{ab}(\sg
)+\Phi^{ab}(\sgm )]\frac{\pr}{\pr \sgm }\dl (\sgm -\sg )+}\non \\
\lf{[\Omega^{ab}(\sg )+\Omega^{ab}(\sgm )]\dl (\sgm -\sg )]+}\non\eeq
\beq\lf{\frac{\pr A}{\pr p_{a}}\frac{\pr B}{\pr p_{b}}[[\omega_{ab}(\sg
)+\omega_{ab}(\sgm )]\ep (\sgm -\sg )+}\non \\
\lf{[\Phi_{ab}(\sg )+\Phi_{ab}(\sgm )]\frac{\pr}{\pr \sgm }\dl (\sgm -\sg
)+}\non \\ \lf{[\Omega_{ab}(\sg )+\Omega_{ab}(\sgm )]\dl (\sgm -\sg
)+}\non\eeq  \beq\lf{[\frac{\pr A}{\pr x^{a}}\frac{\pr B}{\pr
p_{b}}+\frac{\pr A}{\pr p_{b}}\frac{\pr B}{\pr x^{a}}][[\omega_{b}^{a}(\sg
)+\omega_{b}^{a}(\sgm )]\ep (\sgm -\sg )}\non \\ \lf{+[\Phi_{b}^{a}(\sg
)+\Phi_{b}^{a}(\sgm )]\frac{\pr}{\pr \sgm }\dl (\sgm -\sg )]+}\non \\
\lf{[\frac{\pr A}{\pr x^{a}}\frac{\pr B}{\pr p_{b}}-\frac{\pr
A}{\pr p_{b}}\frac{\pr B}{\pr x^{a}}][\Omega _{b}^{a}(\sg )+
\Omega_{b}^{a}(\sgm )]\dl (\sgm -\sg )}\non\eeq The arbitrary functions
$A, B, \omega, \Phi, \Omega$ are functions of $x^{a}(\sg ), p_{a}(\sg )$
and $\omega^{ab}, \omega_{ab}$, $\Phi^{ab},\Phi_{ab}$ are symmetric
functions on $a, b$ and $\Omega^{ab}, \Omega_{ab}$ are antisymmetric
functions.

\subsection{Constant background fields}
 The equations of motion under (PB) $\{,\}_{1}$ are
\beq\lf{\dot x^{a}=g^{ab}[p_{b}-\al B_{bc}x^{'c}]}\\
\lf{\dot p_{a}=\al B_{ab}g^{bc}p_{c}^{'}+[g_{ab}-\al
^{2}B_{ac}g^{cd}B_{db}]x^{''b}}\non\eeq
The bihamiltonity condition is reduce to the following constraints on the
phase space
\beq\label{bih}\lf{\dot
x^{a}=-2\omega_{b}^{a}x^{b}+4\omega^{ab}p_{b}+2\Phi^{ab}p_{b}^{''}-
2\Phi_{b}^{a}x^{''b}+}\non \\
\lf{+2\Omega_{b}^{a}x^{'b}=g^{ab}p_{b}-\al
g^{ab}B_{bc}x^{'c},\,\Omega^{ab}=0}\\
\lf{\dot p_{a}=-4\omega_{ab}x^{b}-2\Phi_{ab}x^{''b}+2\Omega_{ab}x^{'b}+
4\omega_{a}^{b}p_{b}+}\non \\
\lf{+2\Phi_{a}^{b}p_{b}^{''}+2\Omega_{a}^{b}p_{b}^{'}=\al
B_{ab}g^{bc}p_{c}^{'}+}\non \\
\lf{+[g_{ab}-\al ^{2}B_{ac}g^{cd}B_{db}]x^{''b}}\non\eeq
There is the unique solution without constraints
\beq\lf{\omega^{ab}=\frac{1}{4}g^{ab},2\Omega_{b}^{a}=-\al
g^{ac}B_{cb}=\al B_{bc}g^{ca},}\non \\
\lf{-2\Phi_{ab}=g_{ab}-\al^{2}B_{ac}g^{cd}B_{db}}\eeq
The rest structure functions are equal zero. The dual (PB) are
\beq\label{dua}\lf{\{x^{a}(\sg ),x^{b}(\sgm )\}_{2}=\frac{1}{2}g^{ab}\ep
(\sgm -\sg )}\\ \lf{\{x^{a}(\sg ),p_{b}(\sgm )\}_{2}=-\al g^{ac}B_{cb}\dl
(\sgm -\sg )}\non\eeq \beq\lf{\{p_{a}(\sg ),p_{b}(\sgm )\}_{2}=-}\non \\
\lf{-[g_{ab}-\al ^{2}B_{ac}g^{cd}B_{db}]\frac{\pr}{\pr \sgm }\dl (\sgm
-\sg )}\non\eeq
The dual dynamical system is the chiral string
\be\lf{\dot x^{a}=x^{'a},\,\, \dot p_{a}=p_{a}^{'}}\ee
To find the nontrivial dual dynamical system, it needs to introduce some
constraints (\ref{bih}) to the original model. At the beginning, we are
consider the toy model.
\subsection{Relativistic particle in the
background electromagnetic field and noncommutativity}
 The relativistic particle in the constant background electromagnetic field
is described by hamiltonian
\be\lf{H=\frac{1}{2}[(p_{a}+i\beta B_{ab}x_{b})^{2}+m^{2}]}\ee
The electromagnetic field is $A_{a}(x)=-2F_{ab}x_{b}$=$-B_{ab}x_{b}$. The
simplest constraint from (\ref{bih}) without derivatives is
\be\lf{\varphi_{a}=p_{a}+i\al B_{ab}x_{b}\approx 0}\ee
The consistency condition
\be\lf{\{\varphi_{a},H\}=i\al(\al +\beta )B_{ab}\varphi_{b}}\ee
shows, that there are unique set of constraints, if $\al +\beta = 0$. They
are second kind constraints $\{\varphi_{a},\varphi_{b}\}=2i\al B_{ab}$.
There are the following algebra of the phase space coordinates under
the Dirac bracket
\beq\lf{\{x_{a},x_{b}\}_{D}=\frac{-i}{2\al}(B^{-1})_{ab},\{x_{a},p_{b}\}_
{D}=\frac{1}{2}\dl {ab}}\non \\
\lf{\{p_{a},p_{b}\}_{D}=\frac{-i\al}{2}B_{ab}}\eeq
The motion equation under Dirac bracket
\be\lf{\dot x_{a}+2i\al B_{ab}x_{b}=0,\, \dot p_{a}+2i\al
B_{ab}x_{b}=0}\ee
have solution $x_{a}(\tau )=\{e^{-2i\al B\tau}\}_{ab}x_{b}(0)$. The
quantization of the Dirac bracket is reduce to the noncommutative space
\beq\label{reg}\lf{[x_{a},x_{b}]=\frac{1}{2\al}(B^{-1})_{ab},\,[p_{a},p_{b}]=
\frac{\al}{2}B_{ab}}\non \\
\lf{[x_{a},p_{b}]=\frac{i}{2}\dl _{ab}}\eeq
in the analogy to noncommutativity of the open string model \cite{sw}.
But, there is the commutative solution under some restriction on the
background field. We are suppose $\al=-i$ for simplicity and we are
introduce "physical" coordinates
$y_{a}=p_{a}+x_{a}$,$q_{a}=p_{a}-x_{a}$. They form following algebra under
constraint $B_{ac}B_{cb}=\dl_{ab}$
\beq\lf{\{y_{a},y_{b}\}_{D}=0,\{q_{a},q_{b}\}_{D}=0}\non \\
\lf{\{y_{a},q_{b}\}_{D}=-\dl_{ab}-B_{ab}}\eeq
We have after quantization
\beq\lf{[y_{a},y_{b}]=[q_{a},q_{b}]=0,[y_{a},q_{b}]=i(\dl_{ab}-B_{ab})}
\non \\ \lf{q_{a}=-i(\dl_{ab}-B_{ab})\frac{\pr}{\pr y_{a}}}\eeq
As is obvious from (\ref{reg}), the coordinates $x_{a}$ are commute in the
strong coupling regime $\al\to \infty$. The commuting momenta $p_{a}$
must to consider as the coordinates of position in the weak coupling
regime $\al\to 0$.
 The similar condition on the B-field is arise in the closed string model
in the constant background fields with the additional constraint.
\subsection{B-chiral string}
Let me consider following constraint from (\ref{bih})
\be\lf{\varphi_{a}=p_{a}+\beta B_{ab}x^{'b}\approx 0}\ee
The consistency condition
\beq\lf{\{\varphi_{a},\int\limits_{0}^{\pi}d\sgm H_{1}\}_{1}=(\al +\beta )
B_{ab}g^{bc}\varphi_{c}^{'}+}\non \\
\lf{+[g_{ab}-(\al +\beta )^{2}B_{ac}g^{cd}B_{db}]x^{''b}\approx 0}\eeq
If we suppose condition
\be\lf{g_{ab}=(\al +\beta )^{2}B_{ac}g^{cd}B_{db}}\ee
we have first kind constraint $\varphi_{a}$, which is describe the
B-chiral string. The motion equations are
\beq\lf{\dot x^{a}=-(\al +\beta )g^{ab}B_{bc}x^{'c}}\\
\lf{\dot p_{a}=-(\al +\beta )B_{ab}g^{bc}p_{c}^{'},\,\,\ddot
x^{a}=x^{''a}}\non\eeq
This model is the bihamiltonian model under (PB) (\ref{dua}) also. The
B-chiral string model is dual to the chiral model also.
\subsection{Closed string in the arbitrary background fields} The
equations of motion of the closed string, in the arbitrary background
 gravity field and antisymmetric B-field under hamiltonian $H_{1}$
(\ref{ham}) and (PB) $\{,\}_{1}$, are \beq\lf{\dot x^{a}=g^{ab}[p_{b}-\al
B_{bc}x^{'c}]}\\ \lf{\dot p_{a}=\al B_{ab}g^{bc}p_{c}^{'}+[g_{ab}-\al
B_{ac}g^{cd}B_{db}]x^{''b}-}\non \\
\lf{-\frac{1}{2}\frac{\pr g^{bc}}{\pr x^{a}}p_{b}p_{c}-\al\frac{\pr}{\pr
x^{a}}(B_{bd}g^{dc})x^{'b}p_{c}+}\non \\
\lf{+\al\frac{\pr}{\pr x^{b}}(B_{ad}g^{dc})x^{'b}p_{c}-}\non \\
\lf{-\frac{1}{2}\frac{\pr}{\pr
x^{a}}[g_{bc}-\al^{2}B_{bd}g^{de}B_{eb}]x^{'b}x^{'c}+}\non \\
\lf{+\frac{\pr}{\pr
x^{b}}[g_{ac}-\al^{2}B_{ad}g^{de}B_{ec}]x^{'b}x^{'c}}\non\eeq
The bihamiltonity condition for coordinate $x^{a}$ is
\beq\label{arb}\lf{\dot
x^{a}=-2\omega_{b}^{a}x^{b}+4\omega^{ab}p_{b}+2\Phi^{ab}p_{b}^{''}-}\\
\lf{-2\Phi_{b}^{a}x^{''b}+2\Omega_{b}^{a}x^{'b}-2\Omega^{ab}p_{b}^{'}+}
\non \\ \lf{\int\limits_{0}^{\pi}d\sgm [\omega_{b}^{a}x^{'a}+\frac{\pr
\omega^{ac}}{\pr x^{b}}x^{'b}p_{c}+\frac{\pr \omega^{ac}}{\pr
p_{b}}p_{b}^{'}p_{c}]\ep (\sgm -\sg )}\non \\
\lf{+(\frac{\pr \Phi^{ac}}{\pr x^{b}}x^{'b}+\frac{\pr \Phi^{ac}}{\pr p_{b}}
p_{b}^{'})p_{c}^{'}-}\non \\
\lf{-(\frac{\pr \Phi_{c}^{a}}{\pr x^{b}}x^{'b}+\frac{\pr \Phi_{c}^{a}}{\pr
p_{b}} p_{b}^{'})x^{'c}=g^{ab}p_{b}-\al g^{ab}B_{bc}x^{'c}}\non\eeq
The consistency condition of the equation (\ref{arb}) is
\beq\lf{\Phi^{ab}=\Phi_{b}^{a}=\Omega^{ab}=0,\frac{\pr \omega^{ab}}
{\pr p_{c}}=0}\\
\lf{\omega_{b}^{a}=-\frac{\pr \omega^{ac}}{\pr
x^{b}}p_{c},\,2\Omega_{b}^{a}=-\al g^{ac}B_{cb}}\non \\
\lf{\omega^{ab}=Cg^{ab},\,\,\frac{\pr g^{ab}}{\pr x^{c}}x^{c}=ng^{ab},
C=\frac{1}{2(n+2)}}\non\eeq
The metric tensor $g^{ab}(x)$ is homogeneous function of $x^{a}$ order $n$
and $C$ is arbitrary constant.
The bihamiltonity condition for coordinate $p_{a}$ is
\beq\lf{\dot
p_{a}=-2\omega_{ab}x^{b}-2\Phi_{ab}x^{''b}+2\Omega_{ab}x^{'b}+}\\
\lf{+4\omega_{a}^{b}p_{b}+2\Phi_{a}^{b}p_{b}^{''}+2\Omega_{a}^{b}p_{b}^{'}+
\int\limits_{0}^{pi}d\sgm
(\omega_{ab}x^{'b}-}\non \\ \lf{-\frac{\pr^{2}\omega^{cd}}{\pr x^{a}\pr
x^{b}}x^{'b}p_{c}p_{d}-\frac{\pr\omega^{bc}}{\pr x^{a}}p_{b}p_{c}^{'})\ep
(\sgm -\sg )}\non\eeq
\beq\lf{-(\frac{\pr \Phi_{ac}}{\pr x^{b}}x^{'b}+\frac{\pr \Phi_{a}^{c}}{\pr
p_{b}}p_{b}^{'})x^{'c}+}\non \\
\lf{+(\frac{\pr \Phi_{a}^{c}}{\pr x^{b}}x^{'b}+\frac{\pr \Phi_{a}^{c}}
{\pr p_{b}}p_{b}^{'})p_{c}^{'}=}\non\eeq
\beq\lf{=\al
B_{ab}g^{bc}p_{c}^{'}+[g_{ab}-\al^{2}B_{ad}g^{dc}B_{cb}]x^{'b}-}\non \\
\lf{-\frac{1}{2}\frac{\pr g^{bc}}{\pr x^{a}}p_{b}p_{c}-\al\frac{\pr}{\pr
x^{a}}(B_{bd}g^{dc})x^{'b}p_{c}+}\non \\
\lf{+\al \frac{\pr}{\pr x^{b}}(B_{ad}g^{dc})x^{'b}p_{c}-}\non\eeq
\beq\lf{-\frac{1}{2}\frac{\pr}{\pr
x^{a}}[g_{bc}-\al^{2}B_{bd}g^{de}B_{eb}]x^{'b}x^{'c}+}\non \\
\lf{\frac{\pr}{\pr
x^{b}}[g_{ac}-\al^{2}B_{ad}g^{de}B_{ec}]x^{'b}x^{'c}}\non\eeq
The common consistency conditions are
\beq\lf{\Phi^{ab}=0,\Omega^{ab}=0,\Phi_{b}^{a}=0,
\omega^{ab}=Cg^{ab}}\\
\lf{\omega_{ab}=\frac{C}{2}\frac{\pr^{2} g^{cd}}{\pr x^{a}\pr
x^{b}}p_{c}p_{d},\,\,\omega_{b}^{a}=-C\frac{\pr g^{ac}}{\pr
x^{b}}p_{c},}\non \\
\lf{\Phi_{ab}=-\frac{1}{2}[g_{ab}-\al^{2}B_{ac}g^{cd}B_{db}],}\non \\
\lf{\Omega_{ab}=\frac{1}{2}(\frac{\pr \Phi_{bc}}{\pr x^{a}}-\frac{\pr
\Phi_{ac}}{\pr x^{b}})x^{'c}-}\non \\
\lf{-(\frac{\pr \Omega_{b}^{c}}{\pr x^{a}}-\frac{\pr \Omega_{a}^{c}}{\pr
x^{b}})p_{c},}\non \\
\lf{2\Omega_{b}^{a}=-\al g^{ac}B_{ca}, \frac{\pr g^{ab}}{\pr
x^{c}}x^{c}=ng^{ab}, C=\frac{1}{2(n+2)}}\non\eeq
The Jacoby identity
\beq\lf{\{A(\sg ),B(\sgm
)\}_{2}C(\sigma^{''})\}_{2}+}\\ \lf{\{C(\sigma^{''}),A(\sg )\}_{2}B(\sgm
)\}_{2}+}\non \\ \lf{\{B(\sgm ),C(\sigma^{''}\}_{2}A(\sg )\}_{2}=0}\non\eeq
does not be satisfied for arbitrary background fields. The terms with
derivatives of structure functions is reduce to integrability conditions of
the closed string model in the background gravity field and
the antisymmetric B-field. The principal of this conditions is condition
on the structure function $\omega^{ab}(x)$.
\beq\lf{[\frac{\pr g^{ab}(\sg )}{\pr x^{d}}[g^{dc}(\sg )+g^{dc}(\sigma^{''}
)]-}\\ \lf{-\frac{\pr g^{ac}(\sg )}{\pr x^{d}}[g^{db}(\sg
)+g^{db}(\sgm)]]\ep (\sgm -\sg )\ep (\sigma^{''}-\sg )+}\non\eeq
\beq\lf{+[\frac{\pr g^{cb}(\sgm )}{\pr x^{d}}[g^{da}(\sgm )+g^{da}(\sg
)]-}\non \\ \lf{-\frac{\pr g^{ab}(\sgm )}{\pr x^{d}}[g^{dc}(\sgm
)+g^{dc}(\sigma^{''})]]\ep (\sg -\sgm )\ep (\sigma^{''}-\sgm )+}\non\eeq
\beq\lf{[\frac{\pr g^{ac}(\sigma^{''})}{\pr x^{d}}[g^{db}(\sigma^{''}
)+g^{db}(\sgm )]-}\\ \lf{-\frac{\pr g^{cb}(\sigma^{''})}{\pr
x^{d}}[g^{da}(\sigma^{''})+g^{da}(\sg )]]\ep (\sg -\sigma^{''})\ep (\sgm
-\sigma^{''})}\non\eeq It is possible to reduce this condition to unique
equation \beq\lf{[\frac{\pr g^{ab}(\sg )}{\pr x^{d}}[g^{dc}(\sg
)+g^{dc}(\sigma^{''})]-}\\ \lf{\frac{\pr g^{ac}(\sg )}{\pr
x^{d}}[g^{db}(\sg )+g^{db}(\sgm )]]\ep (\sgm -\sg )\ep (\sigma^{''}-\sg
)=0}\non\eeq In assumption, that $g^{ab}= \frac{\pr ^{2}f}{\pr x^{a}\pr
x^{b}}$ , this equation is the matrix equation with the additional
parametric dependence unknown function $f(x^{a}(\sg ))$ of parameter $\sg
$.  The rest equations for structure functions $\omega , \Omega ,\Phi$
together with corresponding generalized functions (\ref{new}) and crossed
terms of the structure functions are similar to previous equation. We do
not reduce tedious expression for this equations here.
\section{Acknowledgements} The author would like to thank A.  Zheltukhin
and D. Sorokin for useful discussions.

\end{document}